\documentclass[twoside,twocolumn]{article}
\usepackage[hmarginratio=1:1,margin={1.45cm,1.45cm}, top=32mm,columnsep=20pt]{geometry} 

\usepackage{amsmath,graphicx,balance,amssymb}
\usepackage{balance}
\def\x{{\mathbf{x}}}

\def\Y{{\mathbf{Y}}}

\def\Yjhat{{\mathbf{\hat{Y}}^{j}}}

\def\Mj{{\mathbf{M}^j}}
\def\Mjtilde{\tilde{\mathbf{M}}^{j}} 

\def\relu{\text{\small{ReLU}}}

\usepackage{microtype} 
\usepackage[english]{babel} 
\usepackage[hang, small,labelfont=bf,up,textfont=it,up]{caption} 
\usepackage{booktabs} 
\usepackage{enumitem} 
\setlist[itemize]{noitemsep} 
\usepackage{balance}
\usepackage{amssymb}
\usepackage{url}
\usepackage{floatrow}
\usepackage{amsmath}
\usepackage{algorithm}
\usepackage{multicol}
\usepackage[noend]{algpseudocode}
\usepackage[normalem]{ulem}
\usepackage{abstract} 
\usepackage{titlesec} 
\renewcommand\thesection{\Roman{section}} 
\renewcommand\thesubsection{\roman{subsection}} 
\titleformat{\section}[block]{\large\scshape\centering}{\thesection.}{1em}{} 
\titleformat{\subsection}[block]{\large}{\thesubsection.}{1em}{} 
\usepackage{fancyhdr} 
\usepackage{titling} 
\usepackage{hyperref} 
\pretitle{\begin{center}\bfseries} 
\posttitle{\end{center}} 
\title{Monaural Singing Voice Separation with Skip-Filtering Connections and Recurrent Inference of Time-Frequency Mask} 
\author{%
	Stylianos Ioannis Mimilakis$^{*}$, Konstantinos Drossos$^{\dagger}$, Jo\~{a}o F. Santos$^{\S\ddagger}$,\\ Gerald Schuller$^{*}$, Tuomas Virtanen$^{\dagger}$, Yoshua Bengio$^{\S\mathparagraph}$ \\
\\
$^*$Fraunhofer IDMT, Ilmenau, Germany\\
	$^{\dagger}$Tampere University of Technology, Tampere, Finland \\
	$^{\S}$Universit\'{e} de Montr\'{e}al, Montreal, Canada \\ 	     	         $^{\ddagger}$INRS-EMT, Montreal, Canada \\
$^{\mathparagraph}$CIFAR Fellow \\
}
\date{} 
\renewcommand{\maketitlehookd}{%
\begin{abstract}
\noindent Singing voice separation based on deep learning relies on the usage of time-frequency masking. In many cases the masking process is not a learnable function or is not encapsulated into the deep learning optimization. Consequently, most of the existing methods rely on a post processing step using the generalized Wiener filtering.
This work proposes a method that learns and optimizes (during training) a source-dependent mask and does not need the aforementioned post processing step. 
We introduce a recurrent inference algorithm, a sparse transformation step to improve the mask generation process, and a learned denoising filter. Obtained results show an increase of 0.49 dB for the signal to distortion ratio and 0.30 dB for the signal to interference ratio, compared to previous state-of-the-art approaches for monaural singing voice separation. 
\end{abstract}
\hspace{0.76cm}\begin{keywords}
Singing voice separation, recurrent encoder decoder, recurrent inference, skip-filtering connections
\end{keywords}
}

\begin{document}

\maketitle

\section{Introduction}
\label{sec:intro}
The problem of music source separation has received a lot attention in the fields of audio signal processing and deep learning~\cite{sisec17}. The most adopted solution is the estimation of a time-varying and source-dependent filter, which is applied to the mixture~\cite{liutkus_alpha}. Performing the filtering operation is done by treating audio signals as wide-sense stationary. This involves transforming the mixture signal using the short-time Fourier transform (STFT). Then, the source-dependent filtering operation is applied to the complex-valued coefficients of the mixture signal. More formally, let $\mathbf{x}$ be the time-domain mixture signal vector of $J$ sources. $\Y \in \mathbb{C}^{M \times N}$ is the complex-valued STFT representation of $\x$, comprising of $M$ overlapping time frames and $N$ frequency sub-bands. The estimation of the $j$-th target source ($\Yjhat \in \mathbb{C}^{M \times N}$) is achieved through:
\begin{equation}\label{eq:masking}
\Yjhat = \Y \odot \Mj \text{,}
\end{equation}
where $\odot$ is element-wise product and $\Mj \in \mathbb{R}_{\geq 0}^{M \times N}$ is the $j$-th source-dependent filter, henceforth denoted as mask.
In~\cite{liutkus_alpha} was shown that a preferred way for estimating the $j$-th source is to derive the mask through the generalized Wiener filtering using $\alpha$-power magnitude spectrograms as:
\begin{equation}\label{eq:themask}
\Mj = \frac{|\Yjhat|^{\circ \alpha}}{\underset{j}{\sum}|\Yjhat|^{\circ \alpha}} \text{,}
\end{equation}
where, $|\cdot|$ and $\circ$ denote the entry-wise absolute and exponentiation operators respectively, and $\alpha$ is an exponent chosen based on the assumed distributions that the sources follow. Finding $\alpha$ (and thus an optimal $\Mj$ for the source estimation process~\cite{liutkus_alpha}) is an open optimization problem~\cite{liutkus_alpha, fitz_masks}.

Deep learning methods for music source separation are trained using synthetically created mixtures $\Y$ (adding signals $\Y^j$ together, i.e., knowing	 the target decomposition). They can be divided into two categories. In the first category, the methods try to predict the mask directly from the mixture magnitude spectrum~\cite{grais16} (i.e. $f_{1}:|\Y| \rightarrow \Mj$). This requires that an optimal $\Mj$ is given (e.g. all the non-linear mixing parameters of the target source are known) during training as a target. However, such information for the $\Mj$ is unknown, and an approximation of $\Mj$ is computed from the training data using Eq.~(\ref{eq:themask}) and empirically chosen $\alpha$ values, under the hypothesis that the source magnitude spectra are additive, which is not true for realistic audio signals ~\cite{liutkus_alpha,fitz_masks}. This implies that such models are optimized to predict non-optimal masks. The methods in the second category try to estimate all sources from the mixture(i.e. $f_{2}:|\Y| \rightarrow |\Yjhat|^{\circ \alpha}\; \forall\, j \in J$) and then use these estimates to compute a mask~\cite{uhl15, cha17, takahashi17, huang, nug16}. This approach is widely adopted, since it is straightforward by employing denoising autoencoders~\cite{bengio_den}, with noise corresponding to the addition of other sources. However, the masks are dependent on the initial $\alpha$-power magnitude estimates of the sources ($|\Yjhat|^{\circ \alpha}$), and the mask computation is not a learned function. Instead, the mask computation uses a deterministic function which takes as inputs the outcomes ($|\Yjhat|^{\circ\alpha} \forall j \in J$) of deep neural networks, e.g. as in~\cite{huang}.

An exception to the above are the works presented in~\cite{mim16} and~\cite{mim17}, where these methods jointly learned and optimized the masking processes described by Eq.~(\ref{eq:masking}) and (\ref{eq:themask}). In~\cite{mim16}, highway networks~\cite{hw15} were shown to be able to approximate a masking process for monaural solo source separation and in~\cite{mim17}, a more robust alternative to~\cite{mim16} is presented. The approach in~\cite{mim17} uses a recurrent encoder-decoder with skip-filtering connections, which allow a source-dependent mask generation process, applicable to monaural singing voice separation. However, the generated masks are not robust against interferences from other music sources, thus requiring a post-processing step using the generalized Wiener filtering~\cite{mim17}.

In this work we present a method for source separation that learns to generate a source-dependent mask which does not require the generalized Wiener filtering as a post-processing step. To do so, we introduce a novel recurrent inference algorithm inspired by~\cite{stoch_depth} and a sparsifying transform~\cite{papyan17} for generating the mask $\Mj$. The recurrent inference allows the proposed method to have a stochastic depth of RNNs during the mask generation process, computing hidden, latent representations which are presumably better for generating the mask. The sparsifying transform is used to approximate the mask using the output of the recurrent inference. In this method the mask prediction is not based on the above mentioned assumptions about the additivity of the magnitude spectrogram of the sources, is part of an optimization process, and is not based on a deterministic function. Additionally, the method incorporates RNNs instead of feed-forward or convolutional layers for the mask prediction. This allows the method to exploit the memory of the RNNs (compared to CNNs) and their efficiency for modeling longer time dependencies of the input data. The rest of the paper is organized as follows: Section~\ref{sec:proposedmethod} presents the proposed method, followed by Section~\ref{sec:experiments} which provides information about the followed experimental procedure. Section~\ref{sec:resndis} presents the obtained results from the experimental procedure and Section~\ref{sec:conc} concludes this work.

\section{Proposed Method}\label{sec:proposedmethod}
\begin{figure}
\includegraphics[width=.9\textwidth]{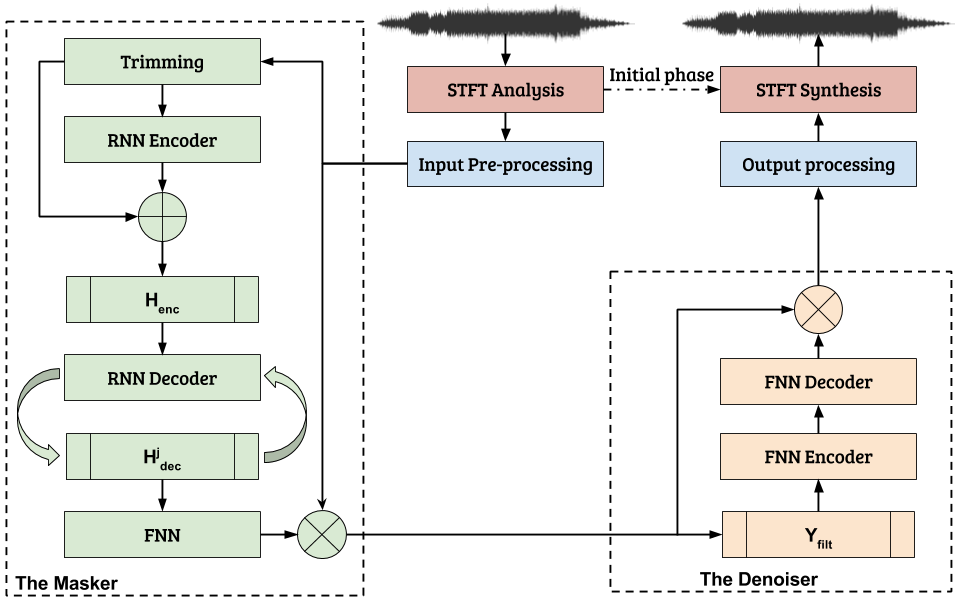}
\caption{Illustration of our proposed method.}
\label{fig:method1}

\end{figure}

Our proposed method takes as an input the time domain samples of the mixture, and outputs time domain samples of the targeted source. The model consists of four parts. The first part implements the analysis and pre-processing of the input. The second part generates and applies a mask, thus creating the first estimate of the magnitude spectrogram of the targeted source. The third part enhances this estimate by learning and applying a denoising filter, and the fourth part constructs the time domain samples of the target source. We call the second part the ``Masker'' and the third the ``Denoiser''. We differentiate between the Masker and the Denoiser because the Masker is optimized to predict a time-frequency mask, whereas the Denoiser enhances the result obtained by time-frequency masking. We implement the Masker using a single layer bi{-}directional RNN encoder ($\text{RNN}_{\text{enc}}$), a single layer RNN decoder ($\text{RNN}_{\text{dec}}$), a feed-forward layer (FFN), and skip-filtering connections between the magnitude spectrogram of the mixture and the output of the FFN. We implement the Denoiser using one FFN encoder ($\text{FFN}_{\text{enc}}$), one FFN decoder ($\text{FFN}_{\text{dec}}$), and skip-filtering connections between the input to the Denoiser and the output of the $\text{FFN}_{\text{dec}}$. We jointly train the Masker and the Denoiser using two criteria based on the generalized Kullback-Leibler divergence ($D_{KL}$), as it is shown in~\cite{fitz_masks, cauchy_nmf} to be a robust criterion for matching magnitude spectrograms. All RNNs are gated recurrent units (GRU). The proposed method is illustrated in Figure~\ref{fig:method1}.

\subsection{Input preprocessing}
Let $\x$ be the vector containing the time-domain samples of a monaural mixture from $J$ sources, sampled at $44.1$kHz. We compute the STFT of $\x$ from time frames of $N=2049$ samples, segmented with Hamming window and a hop size of 384 samples. Each time frame is zero-padded to $N' = 4096$. Subsequent to the STFT we retain only the positive frequencies, corresponding to the first $N=2049$ frequency sub-bands. This yields the complex-valued time-frequency representation of $\x$, $\Y \in \mathbb{C}^{M \times N}$, and the corresponding magnitude $|\Y| \in \mathbb{R}_{\geq 0}^{M \times N}$. We split $|\Y|$ in $B=\lceil M / T \rceil$ subsequences, with $T$ being the length of the subsequence, and $\lceil\cdot\rceil$ is the ceiling function. Each subsequence $b$ overlaps with the preceding one by an empirical factor of $L \times 2$, in order to use some context information for the encoding stage.
We use each subsequence $b$ in $|\Y|$, denoted as $|\Y_\text{in}|$ as an input to the skip-filtering connections (presented later). Furthermore we produce a low-bandwidth version of $|\Y|$, which is used for encoding, by preserving only the first $F = 744$ frequency sub-bands at each frame yielding $|\Y_{\text{tr}}| \in \mathbb{R}_{\geq 0}^{T \times F}$
$|\Y_{\text{tr}}|$.
This operation retains information up to $8$ kHz, in order to reduce the number of trainable parameters but preserving  the most relevant information of the singing voice source.

\subsection{The Masker}

\textbf{RNN encoder}~~We use $|\Y_{\text{tr}}|$ as an input to the $\text{RNN}_{\text{enc}}$. The forward GRU of the $\text{RNN}_{\text{enc}}$ takes $|\Y_{\text{tr}}|$ as an input. The backward one takes as an input the ${|\overleftarrow{\Y}_{\text{tr}}}| = [|\mathbf{y}_{\text{tr}_{T}}|, \ldots, |\mathbf{y}_{\text{tr}_{t}}|, \ldots,$ $|\mathbf{y}_{\text{tr}_{1}}|]\text{,}$ where $|\mathbf{y}_{\text{tr}_{t}}| \in \mathbb{R}_{\geq 0}^{F}$ is a vector in $|\Y_{\text{tr}}|$ at time frame $t$, and $\overleftarrow{ }$ indicates the direction of the sequence. The outputs from the Bi-GRU $\mathbf{h}_{t}$ and $\overleftarrow{\mathbf{h}}_t$ are updated at each time frame $t$ using residual connections~\cite{res_con} and then concatenated as
\begin{equation}
\label{eq:concat}
{\mathbf{h}_{\text{enc}_t}} = \big[({\mathbf{h}}_{t} + |{{\mathbf{y}}_{\text{tr}_{t}}}|)^{\text T}, ({\overleftarrow{\mathbf{h}}_{t}} + {\overleftarrow{|\mathbf{y}}_{\text{tr}_{t}}|})^{\text T}\big]^{\text T}\text{.}
\end{equation}
\noindent
The output of the $\text{RNN}_{\text{enc}}$ for all $t \in T$ is denoted as $\mathbf{H}_{\text{enc}} \in \mathbb{R}^{T \times (2\times F)}$ and is followed by the context information removal defined as:

\begin{equation}
\label{eq:subsam}
\tilde{\mathbf{H}}_{\text{enc}} = [ {\mathbf{h}}_{\text{enc}_{1+L}}, {\mathbf{h}}_{\text{enc}_{2+L}}, 
\ldots, {\mathbf{h}}_{\text{enc}_{T-L}}]\text{, }
\end{equation}
yielding $\tilde{\mathbf{H}}_{\text{enc}} \in \mathbb{R}^{T'\times (2\times F)}$ for $T' = T - (2 \times L)$. Residual connections are used to ease the RNN training~\cite{res_con}.

\noindent
\textbf{Recurrent inference and mask prediction}~~Inspired by recent optimization methods employing stochastic depth~\cite{stoch_depth}, we propose a recurrent inference algorithm that processes the latent variables of the $\text{RNN}_{\text{dec}}$ which affect the mask generation. We use this algorithm in order to employ a stochastic depth for the network parts responsible for predicting the mask, increasing the performance of our method. The recurrent inference is an iterative process and consists in reevaluating the latent variables $\mathbf{H}^{j}_{\text{dec}}$, produced by the $\text{RNN}_{\text{dec}}$, until a convergence criterion is reached, thus avoiding the need to specify a fixed number of applications of the $\text{RNN}_{\text{dec}}$. The stopping criterion is a threshold on the mean-squared-error ($\mathcal{L}_\text{MSE}$) between the consecutive estimates of $\mathbf{H}^{j}_{\text{dec}}$,
with a $\mathcal{L}_\text{MSE}$ threshold $\tau_{\text{term}}$.
A maximum number of iterations ($iter$) is used to 
avoid having infinite iterations for convergence between the above mentioned consecutive estimates.
$\mathbf{H}^{j}_{\text{dec}}$ is used only for the singing voice, i.e. $j=1$. Let $\mathcal{G}^{j}_{\text{dec}}$ be the \emph{source-dependent} and trainable function of the $\text{RNN}_{\text{dec}}$. The recurrent inference is performed using Algorithm~\ref{algo:RI}.

\begin{algorithm}
\small
\begin{algorithmic}[1]
\caption{Recurrent Inference}\label{algo:RI}
\State $\mathbf{S}^{j}_{0} \gets \mathcal{G}^{j}_{\text{dec}}(\tilde{\mathbf{H}}_{\text{enc}})$
	  \For{$i \in \{1,\dots, iter\}$}
      \State$ \mathbf{H}^{j}_{\text{dec}} \gets \mathcal{G}^{j}_{\text{dec}}(\mathbf{S}^{j}_{i - 1})$
      \If{$\mathcal{L}_{\text{MSE}}(\mathbf{S}^{j}_{i-1}, \mathbf{H}^{j}_{\text{dec}}) < \tau_{\text{term}}$}
      \State{Terminate the process}
      \EndIf  
      \State $\mathbf{S}^{j}_{i} \gets \mathbf{H}^{j}_{\text{dec}}$
      \EndFor
      \Return $\mathbf{H}^{j}_{\text{dec}}$
\end{algorithmic}
\end{algorithm}
\normalsize

$\mathbf{H}^{j}_{\text{dec}}$ is then given to the FFN layer with shared weights through time frames, in order to approximate the $j$-th source-dependent mask as:
	\begin{equation}
		\label{eq:maskapprox}
		\Mjtilde = \relu(\mathbf{H}^{j}_{\text{dec}} \mathbf{W}_{\text{mask}} + \mathbf{b}_{\text{mask}})\text{, }
	\end{equation}
\noindent    
where $\relu$ is the element-wise rectified linear unit function producing a sparse~\cite{papyan17} approximation of the target source mask $\Mjtilde \in \mathbb{R}_{\geq 0}^{T' \times N}$. The sparsification is performed in order to improve the interference reduction of~\cite{mim17}.
The ReLU function can produce high positive values inducing distortions to the audio signal. However, the reconstruction loss (see Eq. (\ref{eq:cost})) will alleviate that. $\mathbf{W}_{\text{mask}} \in \mathbb{R}^{(2 \times F) \times N}$ is the weight matrix of the FFN comprising a dimensionality expansion up to $N$, in order to recover the original dimensionality of the data. $\mathbf{b}_{\text{mask}} \in \mathbb{R}^{N}$ is the corresponding bias term.

\noindent
\textbf{Skip filtering connections and first estimate of the targeted source}~~We obtain an estimate of the magnitude spectrum of the target source $|\hat{\mathbf{Y}}_{\text{filt}}^j| \in \mathbb{R}_{\geq 0}^{T' \times N}$ through the skip-filtering connections as:
\begin{align}\label{eq:sfilt}
|\hat{\Y}_{\text{filt}}^j| =& |\tilde{\Y}_{\text{in}}| \odot \Mjtilde \text{, where}\\
|\Y_{\text{in}}| =& 
[|\mathbf{y}_{\text{in}_{L}}|, \cdots, |\mathbf{y}_{\text{in}_{T-L}}|]\text{.}
\end{align}
\subsection{The Denoiser}
The output of the Masker is likely to contain interferences from other sources \cite{mim17}. The Denoiser aims to learn a denoising filter for enhancing the magnitude spectrogram estimated by this masking procedure. This denoising filter is implemented by an encoder-decoder architecture with the $\text{FFN}_{\text{enc}}$ and $\text{FFN}_{\text{dec}}$ of Fig.~\ref{fig:method1}. $\text{FFN}_{\text{enc}}$ and $\text{FFN}_{\text{dec}}$ have shared weights through time frames. The final enhanced magnitude spectrogram estimate of the target source $|\hat{\Y}^j|$ is computed using

	\begin{equation}
		\label{eq:senhancment}
		|\hat{\Y}^j| = \relu(\relu(|\hat{\Y}_{\text{filt}}^j| \mathbf{W}_{\text{enc}} + \mathbf{b}_{\text{enc}}) \mathbf{W}_{\text{dec}} + \mathbf{b}_\text{dec}) \odot |\hat{\Y}_{\text{filt}}^j| \text{,}
	\end{equation}
\noindent where $\mathbf{W}_{\text{enc}} \in \mathbb{R}^{N \times \lfloor N/2 \rfloor}$ and $\mathbf{W}_{\text{dec}} \in \mathbb{R}^{\lfloor N/2 \rfloor \times N}$ are the weight matrices of the $\text{FFN}_{\text{enc}}$ and $\text{FFN}_{\text{dec}}$, with the corresponding biases $\mathbf{b}_{\text{enc}} \in \mathbb{R}^{\lfloor N/2 \rfloor}$, $\mathbf{b}_{\text{dec}} \in \mathbb{R}^{N}$, respectively. $\lfloor \cdot \rfloor$ denotes the floor function.
\subsection{Training details and post-processing}
We train our method to minimize the objective consisting of a reconstruction and a regularization part as:
\begin{equation}
		\label{eq:cost}
		\begin{split}
			\mathcal{L} = D_{KL}(|{\Y}^j|\;||\;|\hat{\Y}^j|) + \lambda_{\text{rec}}D_{KL}(|{\Y}^j|\; || \; |\hat{\Y}_{\text{filt}}^j|) \\ + \lambda_{mask}|\text{diag}\{\mathbf{W}_{\text{mask}}\}|_1 + \lambda_{\text{dec}}||\mathbf{W}_{\text{dec}}||_2^2 \text{,}
		\end{split}
	\end{equation}
	where $|{\Y}^j|$ is the magnitude spectrogram of the true source, diag$\{\cdot\}$ denotes the elements on the main diagonal of a matrix, $|\cdot|_1$, and $||\cdot||_2^2$ are the $\ell_1$ vector norm and the squared matrix $L_2$ norm respectively, and $\lambda_{\text{mask}}$, and $\lambda_{\text{dec}}$ are scalars. For $\lambda_{\text{rec}}$ the following condition applies:
	\begin{equation}
		\label{eq:lrec}
		\lambda_{\text{rec}} = 
		\begin{cases}
			1, & \text{if}\ D_{KL}(|{\Y}^j| \; || \; |\hat{\Y}_{\text{filt}}^j|) \geq \tau_{\text{rec}}\\ & \text{and } D_{KL}(|{\Y}^j|\;||\;|\hat{\Y}^j|) \geq \tau_{\text{min}} \\
			0, & \text{otherwise}
		\end{cases}\text{,}
	\end{equation}
	where $\tau_{\text{rec}}$ and $\tau_{\text{min}}$ are hyper-parameters penalizing the mask generation process, allowing a collaborative minimization of the overall objective. The usage of $\lambda_{\text{rec}}D_{KL}(|{\Y}^j|\; || \; |\hat{\Y}_{\text{filt}}^j|)$
 will ensure that $\Mj$ can be used to initially estimate the target source, which is then improved by employing the Denoiser and $D_{KL}(|{\Y}^j|\;||\;|\hat{\Y}^j|)$.   
 The penalization of the elements in the main diagonal of $\mathbf{W}_{\text{mask}}$ will ensure that the generated mask is not something trivial (e.g. a voice activity detector), while the reconstruction losses using the $D_{KL}$ will ensure that a source-dependent mask is generated, that minimizes the aforementioned distance. The squared matrix $L_2$ norm is employed to improve the generalization of the model.

By processing each subsequence using the proposed method, the estimates are concatenated together to form $|\hat{\Y}^j| \in \mathbb{R}_{\geq 0}^{M\times N}$. For the singing voice we retrieve the complex-valued STFT $\hat{Y}^{j=1}$ by means of 10 iterations of the Griffin-Lim algorithm (least squares error estimation from modified STFT magnitude)~\cite{gla} initialized with the mixture's phase and using $|\hat{\Y}^j|$. The time-domain samples $\mathbf{\hat{x}}^{j=1}$ are obtained using inverse STFT.
\section{Experimental Procedure}
\label{sec:experiments}
We use the development subset of Demixing Secret Dataset (DSD$100$)\footnote{\url{http://www.sisec17.audiolabs-erlangen.de}} and the non-bleeding/non-instrumental stems of MedleydB~\cite{medleydb} for the training and validation of the proposed method. The evaluation subset of DSD$100$ is used for testing the objective performance of our method. For each multi-track contained in the audio corpus, a monaural version of each of the four sources is generated by averaging the two available channels. For training, the true source $|\mathbf{Y}^j|$ is the outcome of the ideal ratio masking process~\cite{ps_masks}, element-wise multiplied by a factor of $2$. This is performed to avoid the inconsistencies in time delays and mixing gains between the mixture signal and the singing voice (apparent in MedleydB dataset). The length of the sequences is set to $T = 60$, modeling approximately $0.5$ seconds, and $L = 10$. The thresholds for the minimization of Eq.(\ref{eq:cost}) are $\tau_{\text{rec}} = 1.5$ and $\tau_{\text{min}} = 0.25$ and the corresponding scalars are $\lambda_{\text{mask}} = 1e^{-2}$, and $\lambda_{\text{dec}} = 1e^{-4}$.
The hidden to hidden matrices of all RNNs were initialized using orthogonal initialization~\cite{saxe} and all other matrices using Glorot normal~\cite{glorot}.
All parameters are jointly optimized using the Adam algorithm~\cite{adam}, with a learning rate of $1e^{-4}$, over batches of $16$, an $L_2$ based gradient norm clipping equal to $0.5$ and a total number of $100$ epochs. All of the reported parameters were chosen experimentally with two random audio files drawn from the development subset of DSD$100$. The implementation is based on PyTorch\footnote{\url {http://pytorch.org/}}.

We compared our method with other state-of-the-art approaches dealing with monaural singing voice separation, following the standard metrics, namely signal to noise ratio (SIR) and signal to distortion ratio (SDR) expressed in dB, and the rules proposed in the music source separation evaluation campaign~\cite{sisec17} (e.g. using the proposed toolbox for SIR and SDR calculation). The compared methods are: i) GRA: Deep FFNs~\cite{grais16} for predicting both binary and soft masks~\cite{ps_masks} which are then combined to provide source estimates, ii) CHA: A convolutional encoder-decoder for magnitude source estimation, without a trainable mask approximation~\cite{cha17} iii) MIM-HW: Deep highway networks for music source separation~\cite{mim16} approximating the filtering process of Eq.(1), retrained using the development subset of DSD$100$, and iv) MIM-DWF, MIM-DWF$^{+}$: The two GRU encoder-decoder models combined with generalized Wiener filtering~\cite{mim17}, trained on the development subset of DSD$100$ (MIM-GRUDWF) and the additional stems of MedleydB (MIM-DWF$^{+}$). The methods denoted as MIM-HW, MIM-DWF, and MIM-DWF$^{+}$ were re-implemented for the purposes of this work. For the rest of the methods we used their reported evaluation results obtained from~\cite{sisec17}. Our proposed methods are denoted as \underline{GRU-NRI}, which does not include the recurrent inference algorithm, and two methods using different hyper-parameters for the recurrent inference algorithm: \underline{GRU-RIS}$^{\text{s}}$, parametrized using a maximum number of iterations $iter = 3$, and $\tau_{\text{term}} = 1e^{-2}$, and \underline{GRU-RIS}$^{\text{l}}$ parametrized using a maximum number of iterations $iter = 10$ and $\tau_{\text{term}} = 1e^{-3}$, which where selected according to their performance in minimizing~Eq. (\ref{eq:cost}).
\section{Results \& Discussion}\label{sec:resndis}
Table 1 summarizes the results of the objective evaluation for the aforementioned methods by showing the median values obtained from the SDR and SIR metrics. The proposed method based on recurrent inference and sparsifying transform is able to provide state-of-the-art results for monaural singing voice separation, without the necessity of post-processing steps such as generalized Wiener filtering, and/or additionally trained deep neural networks. Compared to methods that approximate the masking processes (GRA, MIM-HW, MIM-DWF, and MIM-DWF$^{+}$) there are significant improvements in overall median performance of both the SDR and SIR metrics, especially when the masks are not a learned function, such as in the case of CHA. 
\begin{table}
	\label{table:results}
    \centering
    \caption{\small Median SDR and SIR values in dB for the investigated approaches. Proposed approaches are underlined. Higher is better.}
\scalebox{0.85}{
\begin{tabular}{llllll}
\textbf{Method} & \textbf{SDR} & \multicolumn{1}{l|}{\textbf{SIR}} & \textbf{Method} & \textbf{SDR} & \textbf{SIR} \\ \hline
GRA\cite{grais16} & -1.75 & \multicolumn{1}{l|}{1.28} & MIM-DWF\textsuperscript{+}\cite{mim17} & 3.71 & 8.01 \\
MIM-HW\cite{mim16} & 1.49 & \multicolumn{1}{l|}{7.73} & {\underline{GRU-NRI}} & 3.62 & 7.06 \\
CHA\cite{cha17} & 1.59 & \multicolumn{1}{l|}{5.20} & {\underline{GRU-RIS\textsuperscript{s}}} & 3.41 & {\underline{\textbf{8.32}}} \\
MIM-DWF\cite{mim17} & 3.66 & \multicolumn{1}{l|}{8.02} & {\underline{GRU-RIS\textsuperscript{l}}} & {\underline{\textbf{4.20}}} & 7.94
\end{tabular}
    }
\end{table}
Using the proposed method, a gain of $0.49$ dB for the SDR is observed between MIM-DWF$^{+}$ and GRU-RIS$^\text{l}$ and $0.30$ dB for the SIR between the MIM-DWF and GRU-RIS$^\text{s}$. Finally, by allowing a larger number of iterations during the recursive inference the mask generation performance and using skip-filtering connections we see an increase in SDR which outperforms the previous methods MIM-DWF and MIM-DWF$^{+}$, but at the cost of a loss in SIR. A demo for the proposed method is available at {\url{https://js-mim.github.io/mss_pytorch/}}.
\section{Conclusion}
\label{sec:conc}
In this work we presented an approach for singing voice separation that does not require post-processing using generalized Wiener filtering. We introduced to the skip-filtering connections~\cite{mim17} a sparsifying transform yielding comparable results to approaches that rely on generalized Wiener filtering. Furthermore, the introduced recurrent inference algorithm was shown to provide state-of-the-art results in monaural singing voice separation. Experimental results show that these extensions outperform previous deep learning based approaches for singing voice separation.
\section{Acknowledgements}
\label{sec:acknowledgements}
The research leading to these results has received funding from the European Union's H2020 Framework Programme (H2020-MSCA-ITN-2014) under grant agreement no 642685 MacSeNet. Part of the computations leading to these results were performed on a TITAN-X GPU donated to K. Drossos from NVIDIA. The authors would like to thank Paul Magron for the precious feedback.

\end{document}